# NMR GHZ

R. Laflamme[1], E. Knill[2], W.H. Zurek[1], P. Catasti[3], S.V.S. Mariappan[3]

[1] *Theoretical Astrophysics T-6, MS B-288;*
[2] *Computer Research and Applications CIC-3, MS B-265;*
[3] *Theoretical Biology T-10, MS B-288*
*Los Alamos National Laboratory, Los Alamos, NM 87455*
(March 8, 2018)

We describe the creation of a Greenberger-Horne-Zeilinger (GHZ) state of the form $(|000\rangle + |111\rangle)/\sqrt{2}$ (three maximally entangled quantum bits) using Nuclear Magnetic Resonance (NMR). We have successfully carried out the experiment using the proton and carbon spins of trichloroethylene, and confirmed the result using state tomography. We have thus extended the space of entangled quantum states explored systematically to three quantum bits, an essential step for quantum computation.

We live in a world which, to the best of our knowledge, is surprisingly well described by the laws of quantum mechanics. Fundamental processes in nature are all compatible with quantum mechanics, even if some of its predictions are counterintuitive.

A good example of this peculiar behavior is given by the now famous pairs of two state systems described by Einstein, Podolosky and Rosen (EPR) [1]. The unorthodox behavior of this composite system has been crystallized by the Bell inequalities [2], which give a statistical test for the existence of *elements of reality* [3]. In our world, photons in a polarization state of the form $(|00\rangle - |11\rangle)/\sqrt{2}$ violate these inequalities as shown by Aspect et al. [4] and therefore contradict the existence of elements of reality.

In a beautiful paper, Greenberger, Horne and Zeilinger (GHZ) [5] demonstrated that it is possible, in a single run of experiments using a state of the form $|000\rangle + |111\rangle$ (the GHZ state), to refute the existence of elements of reality. The GHZ experiment has been succinctly summarized by Mermin [6]. Even though this experiment is very appealing, nobody has yet been able to perform it. The reason is that it is rather difficult to precisely manipulate entangled quantum states of many particles. In fact, up to now, entangled pure states of only two particles have been systematically explored [4,7,8,9].

In this letter we show how three particle entangled states can be realized using nuclear magnetic resonance techniques. We have carried out the experiment and verified that indeed we had a GHZ state by using tomography [10]. We first describe how it is possible to obtain a pure state result from the initial mixed density matrix of a NMR system. Then we explain the sequence of operations used to obtain a GHZ state and finally we give the experimental results.

We will not investigate the non-local behavior of GHZ states, as NMR is not appropriate for this undertaking. What we have created, however, is a typical state needed for a three-bit quantum computer. Indeed our approach is the one used for quantum computation and we refer the reader to [11] for an introduction.

The usual approach for quantum computation requires an initial pure state [11]. The computation itself is a transformation of this initial state using unitary operations. Useful information is then extracted by a measurement of the final state. However, it has recently been shown that the computation can be performed using an initial mixed state as long as the decoherence time is sufficiently long [12,13,14,15].

In liquid state NMR, the computation takes place on a large ensemble of identical quantum systems. Each member of the ensemble of quantum systems consists of the interacting nuclear spins of a molecule in a high magnetic field. The initial state of the nuclear spins is achieved by allowing the system to relax to thermal equilibrium. Information processing with such an ensemble can be divided into three steps consisting of preparation, computation and readout. Each of these steps is equivalent to an application of certain quantum operations identically to each member of the ensemble.

The nuclear spins are manipulated by applying radio frequency (RF) pulses tuned to the Larmor frequencies of the spins [16]. The spins can be selectively excited by exploiting differences in Larmor frequencies. Entanglement and two spin operations are achieved by a delay to allow for interaction between spins. In our case, these interactions are scalar couplings which can be selectively turned off by the use of refocusing pulses tuned to one of the nuclei. Any unitary quantum operation can be decomposed into such operations [17,18,19].

The measurement step in NMR consists of observing the signal induced in RF coils by the precession of the nuclear spins [16]. In effect, we measure the time evolution of the expectation of the $\sigma_x$ and $\sigma_y$ operators for each nuclear spin. By Fourier transforming the signal and analyzing the spectrum, other operators such as tensor products of either $\sigma_x$ or $\sigma_y$ with $I$ or $\sigma_z$ for pairs of interacting spins can be measured. The traceless part of the density matrix (called the deviation matrix) can be determined by a tomography procedure [10]. This involves several measurements of spectra after applying different reading pulses to the final state, thus permitting observation of all elements of the deviation matrix.

The main problem in using NMR for the observation of multiparticle entanglements and quantum computation is that the sample is initially in a highly mixed state. There are several methods for overcoming this limitation without cooling the sample. One idea is to transform the initial mixed state so that we have a *pseudo-pure* state

$$\rho_{pp} = p \left|0\ldots0\right\rangle\left\langle0\ldots0\right| + \frac{1-p}{N}I, \qquad (1)$$

where $N$ is $2^n$ for $n$ qubits and $p+(1-p)/N$ is the probability of the ground state ($\left|0\ldots\right\rangle$). In NMR this state is indistinguishable from a pure state because all observables are traceless. The methods discussed in the literature [12,13,14,15] are all based on the idea of viewing the non-ground state components of the initial thermal density matrix as noise and applying an averaging method to eliminate their contribution to measurements. These methods are in principle sufficient for exploiting the advantages of quantum computing in an ensemble setting. In our experiments we used a new method for extracting pseudo-pure states from experimental data. The method is based on two features of our experiment. The first is that the sequence of pulses and delays applied to the sample for preparing the GHZ state are much shorter than the decoherence time for the nuclear spins (about 5ms versus at least 210ms). As a result, the evolution is very close to unitary and therefore preserves the eigenvalue structure of the initial state. This structure is well known and determined by the thermal distribution. The unitarity property can be tested by comparing the eigenvalues of the output state to those of the thermal distribution. The second feature is that complete state tomography is used to analyze the output of the experiment. This allows us to decompose the output state into its eigenstates. The state $\left|000\right\rangle$ is associated with the smallest eigenvalue of the thermal state and must be transformed by unitary evolution into the corresponding eigenstate of the output. By comparing this eigenstate to the desired GHZ state, we learn whether our pulses indeed generated this state from that component of the ensemble which was initially in $\left|000\right\rangle$.

In order to create a GHZ state we need three nuclei. A convenient system is trichloroethylene, the molecule shown in Figure 1. The spins of the hydrogen and carbon nuclei were used for the three quantum bits. They interact only weakly with the chlorine nuclei, which can therefore be ignored. In a strong static magnetic field (11.5 Tesla along the z-axis) the evolution of the hydrogen and the two carbon nuclei are well described by the Hamiltonian

$$H = -\omega_H \sigma_z^H - \omega_{C_1}\sigma_z^{C_1} - \omega_{C_2}\sigma_z^{C_2} + J_{HC_1}\sigma_z^H\sigma_z^{C_1} +$$
$$J_{C_1C_2}(\sigma_x^{C_1}\sigma_x^{C_2} + \sigma_y^{C_1}\sigma_y^{C_2} + \sigma_z^{C_1}\sigma_z^{C_2}) + J_{HC_2}\sigma_z^H\sigma_z^{C_2} \quad (2)$$

with $\omega_H \approx 500.1334915$ MHz, $\omega_{C_1} \approx 125.7725805$ MHz, $\omega_{C_2} \approx 125.7732305$ MHz, which gives a chemical shift of 650 Hz between the carbons. The $J$ couplings have values of $J_{HC_1} \approx 203$ Hz, $J_{C_1C_2} \approx 102$ Hz and $J_{HC_2} \approx 10$ Hz.

The data was acquired with a Bruker DRX-500 spectrometer using doubly labeled trichloroethylene ($^{13}C_1, ^{13}C_2, 99\%$). The relaxation time ($T_1$) for the hydrogen is $7s$ and $30s$ for the carbons. The phase decoherence time ($T_2$) for hydrogen is $3s$ and $0.4s$ and $0.2s$ for $C_1$ and $C_2$ respectively.

A simple circuit to create a GHZ state consists of a rotation by $\pi/2$ around the y-axis followed by two CONTROL-NOT gates on the other qubits. Adapting this to our system results in the pulse sequence shown in figure 2 b). The deviation density matrix, initially given by the state

$$\rho_\Delta^i = \omega_H \sigma_z^h \otimes 1 \otimes 1 + \omega_{C_1} 1 \otimes \sigma_z^{C_1} \otimes 1 + \omega_{C_2} 1 \otimes 1 \otimes \sigma_z^{C_2} \quad (3)$$

is transformed to

$$\rho_\Delta^f = -\omega_H \sigma_z^h \otimes \sigma_z^{C_1} \otimes 1 - \omega_{C_1} \sigma_x^h \otimes \sigma_x^{C_1} \otimes 1$$
$$-\omega_{C_2} 1 \otimes \sigma_z^{C_1} \otimes \sigma_z^{C_2} \quad (4)$$

The tomography procedure was implemented with twelve sets of reading pulses. For each set, two experiments were performed to read off the hydrogen and the carbon spectra.

To compute the peak intensities for each spectrum, we assumed that each peak is approximately Lorentzian. The peak positions and decay parameters associated with the Lorentzian shape were obtained by optimizing (in the least squares sense) the match to the calibration spectra. The carbon and hydrogen spectra were both matched to within 2%. This should be compared to the estimated noise, which was determined to be less than .5% for the carbon and .01% for the hydrogen nuclei. The mismatch is due to additional peaks, primarily from unlabeled compound, and also to shimming problems causing deviation from the ideal Lorentzian lineshape. The decay parameters (essentially $T_2^*$) obtained after optimization were $0.51 \pm 0.03$sec for the hydrogen nucleus, $0.41 \pm 0.01$sec for carbon 1 and $0.23 \pm 0.01$sec for carbon 2 (the error bars are estimated from the variation between different peaks of the same nucleus). The latter agree well with the experimentally determined $T_2$ time for the carbons. Since $T_2$ for the hydrogen is near 3sec, it can be seen that substantial peak broadening due to magnetic field inhomogeneity reduces the hydrogen $T_2^*$.

After the peak positions and decay parameters were determined, each spectrum of the experiment was deconvolved as a linear combination of the ideal peaks. The coefficients of the combination yield the intensity and phase of each peak in the spectrum. These numbers were then normalized by the calibration intensities and phase corrected using the calibration phases. Then they were used to determine the deviation density matrix of the output state of the experiment.

An idea of how *unitary* the evolution has been is obtained by comparing the eigenvalues of the initial deviation matrix ($\{24, 16, 16, 8, -8, -16, -16, -24\}$) to the final one ($\{24.1, 16.3, 15.5, 7.7, -8.1, -15.0, -16.5, -23.8\}$). Clearly the transformation was unitary to a very good approximation.

Using the procedure explained above we get, after diagonalizing and taking the appropriate eigenvector, the experimentally determined GHZ state

$$\rho_e^{GHZ} = 10^{-3} \begin{bmatrix} 493 & -53 & -47-18i & -28-5i & 25-22i & -14+31i & -31-13i & 468-147i \\ -53 & 6 & 5+2i & 3 & -3+2i & 2-3i & 3+i & -51+16i \\ -47+18i & 5-2i & 5 & 3 & -2+3i & -3i & 3 & -39+31i \\ -28+5i & 3 & 3 & 2 & -1+2i & -2i & 2 & -25+13i \\ 25+22i & -3-2i & -2-3i & -1-2i & 2 & -2 & -2i & 30+14i \\ -14-31i & 2+3i & 3i & & -2 & 2 & 2i & -23-25i \\ -31+13i & 3-i & 3 & 2 & 2i & -2i & 2 & -25+21i \\ 468+147i & -51-16i & -39-31i & -25-13i & 30-14i & -23+25i & -25-21i & 488 \end{bmatrix}. \quad (5)$$

A pictorial representation is given in figure 3. The fidelity of the state compared to the ideal GHZ state is

$$\mathcal{F} = \langle \Psi_{GHZ} | \rho_e^{GHZ} | \Psi_{GHZ} \rangle = 0.95. \quad (6)$$

In conclusion, we have shown how to construct an effective GHZ state with a fidelity of 95% in NMR starting with the thermal mixed state. The experiments demonstrate the ability in NMR to fully explore the state space of multi-particle systems, which is all that is required for quantum computation. By using a four spin system, the paradoxical output of the originally proposed GHZ experiments can be observed as proposed in [20]. However, due to the microscopic separation of the particles involved and the method for observation used, this would not be a true test of the existence of elements of reality.

To construct our GHZ state we developed a new method for extracting pseudo-pure states from NMR spectra. The method can be used efficiently for process tomography [10], since each experiment determines the transformation of the state space induced by the applied operation on each of the eigenstates of the input state.

Our experiments demonstrate that room temperature liquid NMR is well suited for quantum computations involving small numbers of qubits. Although manipulating three qubits is a small step for large scale quantum computation, it is the first time that a quantum network has been used to systematically entangle more than two qubits.

**Acknowledgments.** We would like to thank David Cory and Timothy Havel for useful conversations and the Stable Isotope Laboratory at Los Alamos for the use of their facility. This research was supported in part by the National Security Agency.

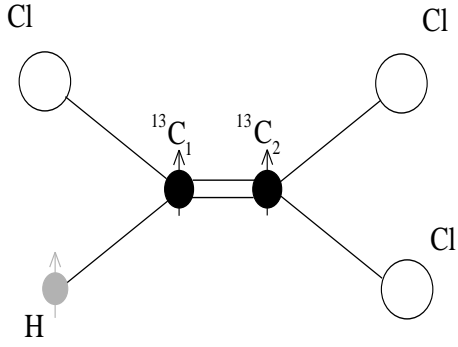

FIG. 1. Pictorial representation of trichloroethylene. The three qubits are given by the nuclei of hydrogen ($H$) and of the two carbon 13's ($C_1$ and $C_2$). The last two are distinguishable because of the assymmetry of the chlorine environment. The chlorine nuclei have spin 3/2 and their interaction with the qubits can be neglected.

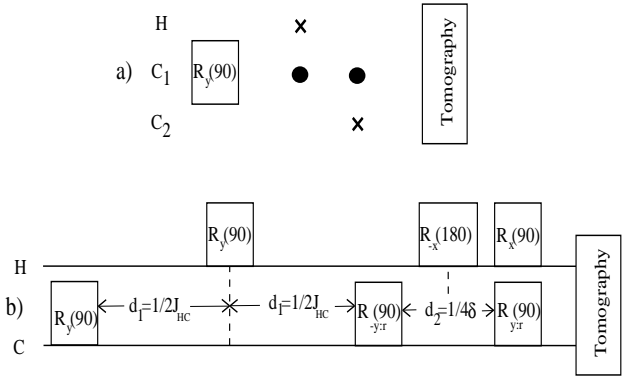

FIG. 2. a) Circuit to create a GHZ state. $R_a(\theta)$ corresponds to a rotation of the qubit by an angle $\theta$ around the $a$-axis and the two other gates are CONTROL-NOTs (see DiVincenzo[14]). b) Implementation by an NMR pulse sequence producing a GHZ state (in the rotating frame of $H$ and $C_1$). The delay $d_1$ serves to let the coupling between the nuclei create the engtanglement, and $d_2$ is used to generate a phase shift on the $C_2$ nucleus. Coupling between the carbon nuclei during this delay is negligible. All pulses are non-selective.

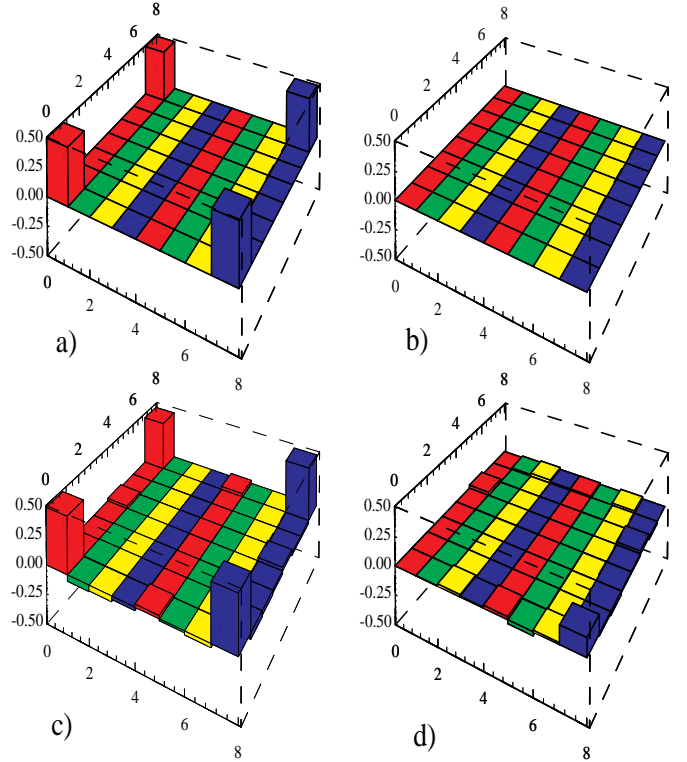

FIG. 3. Pictorial representations of the theoretical and experimental density matrices $\rho$ for the GHZ state $|000\rangle + |111\rangle$. a), b) Real and imaginary parts of the theoretically determined density matrix. c), d) Real and imaginary parts of the experimentally measured density matrix. Each bar graph represents the value of the transitions between the eight states $|000\rangle \ldots |111\rangle$ by the height of the bar in the corresponding position of the $8 \times 8$ array.

4